# Flashmon V2: Monitoring Raw NAND Flash Memory I/O Requests on Embedded Linux

Pierre Olivier
Univ. Europeenne de Bretagne
Univ. Bretagne Occidentale,
UMR6285, Lab-STICC,
F29200 Brest, France,
pierre.olivier@univ-brest.fr

Jalil Boukhobza
Univ. Europeenne de Bretagne
Univ. Bretagne Occidentale,
UMR6285, Lab-STICC,
F29200 Brest, France,
jalil.boukhobza@univ-brest.fr

Eric Senn
Univ. Europeenne de Bretagne
Univ. Bretagne Sud,
UMR6285, Lab-STICC,
F56100 Lorient, France
eric.senn@univ-ubs.fr

## ABSTRACT
This paper presents Flashmon version 2, a tool for monitoring embedded Linux NAND flash memory I/O requests. It is designed for embedded boards based devices containing raw flash chips. Flashmon is a kernel module and stands for "flash monitor". It traces flash I/O by placing kernel probes at the NAND driver level. It allows tracing at runtime the 3 main flash operations: page reads / writes and block erasures. Flashmon is (1) generic as it was successfully tested on the three most widely used flash file systems that are JFFS2, UBIFS and YAFFS, and several NAND chip models. Moreover, it is (2) non intrusive, (3) has a controllable memory footprint, and (4) exhibits a low overhead (< 6%) on the traced system. Finally, it is (5) simple to integrate and used as a standalone module or as a built-in function / module in existing kernel sources. Monitoring flash memory operations allows a better understanding of existing flash management systems by studying and analyzing their behavior. Moreover it is useful in development phase for prototyping and validating new solutions.

## Categories and Subject Descriptors
D.4.2 [**Operating Systems**]: Storage Management – *secondary storage;* D.4.8 [**Operating Systems**]: Performance – *monitors;* D.4.8 [**Memory Structures**]: Semiconductor Memories

## 1. INTRODUCTION
NAND flash memory is the main secondary storage media in embedded systems. This is due to the many benefits it provides: small size, shock resistance, low power consumption, and I/O performance. According to Forward Insight [10], the embedded NAND flash market is predicted to reach nearly 40 000 millions GB in 2014, which is more than 57% of the total NAND flash market. NAND flash presents some specific constraints that require the implementation of dedicated and complex management mechanisms. One of these mechanisms is implemented by the Operating System (OS) in the form of dedicated Flash File Systems (FFS). That solution is adopted in devices using raw flash chips on embedded boards, such as smartphones, tablet PCs, set-top boxes, etc. Linux is a major operating system in such devices, and provides a wide support for several NAND flash memory models. In these devices the flash chip itself does not embed any particular controller.

This paper presents Flashmon version 2, a tool for monitoring embedded Linux I/O operations on NAND secondary storage. Monitoring such operations is essential to (1) study the behavior of flash management mechanisms, (2) evaluate their performance, and (3) helps in proposing optimizations. It allows to supplement storage performance studies that are mainly based on execution time measurements, by giving details regarding the low level flash operations to explain performance behaviors. It also helps in prototyping and validating new flash management systems. Flashmon 2 is the successor of Flashmon 1 [2].

To the best of our knowledge, there is no existing tool allowing to monitor raw NAND chips operations in embedded systems. One of the tools which is the closest to Flashmon is *Blktrace* [4]. Blktrace monitors operations at the block layer level: I/O requests queuing, merging, completion, etc. It is thus designed for block devices, primarily hard disks.

This paper is organized as follows: in section 2, general concepts on NAND flash memory and its management with Linux are presented. Flashmon version 2 implementation and features are depicted in section 3. In section 4, case studies are presented as examples of Flashmon usage interest. An analysis of Flashmon's impact on the traced system is given in section 5, before concluding in section 6.

## 2. NAND FLASH MEMORY
### 2.1 General Concepts on Flash Memory
Flash memory is a non volatile memory based on floating gate transistors [3, 11]. NAND flash subtype is block addressed and offers a high storage density: it is dedicated to data storage. A NAND flash chip architecture is organized in a hierarchical way [5]. A chip is composed of one or more dies. A die contains one or more planes. A plane contains a matrix of blocks, which are composed of pages. NAND flash supports three main operations: *read* and *write* operations are achieved at the page level, and the *erase* operation is performed on a whole block. For the sake of simplicity, in this paper we consider a flash chip as a set of blocks containing pages. Typically, the size of today's NAND flash pages is 2, 4 or 8 KB. Blocks contain a number of pages that is multiple of 32 (typically 64). Average operations latencies are between 25 and 200 μs for reads, between 250 and 500 μs for writes, and up to 2 ms for the erase operation [11].

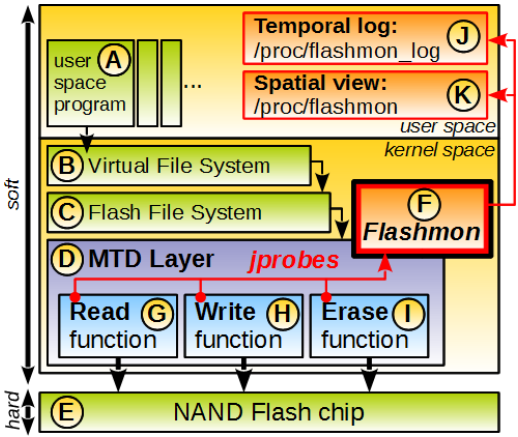

**Figure 1: Linux software involved in NAND flash storage management, and Flashmon integration in this stack**

NAND flash presents specific constraints. The first is 1) the *erase-before-write* rule, which states that a page already containing data cannot be overwritten and must first be erased. Because the target of the erase operation is a whole block, flash management mechanisms perform out-of-place data updates. Old data versions are not erased synchronously but rather invalidated. A process named the garbage collector is in charge of recycling (erasing) invalid data to create free space. Another constraint is 2) the fact that a flash block can only sustain a limited number of erase operations, after which it can no longer retain data. Wear leveling techniques are implemented by flash management mechanisms to distribute evenly the write and erase cycles over the whole flash array to maximize the flash memory lifetime. Blocks that are worn out, called bad blocks, must be coped with. In addition to the above-mentioned constraints, reading / writing the flash array may cause some disturbance leading to random bitflips. Flash management systems must implement Error-Correcting-Codes (ECC) algorithms. To avoid such disturbance, writes inside a flash block must be sequential.

## 2.2 Embedded Linux Raw NAND Chips Management

Linux manages raw NAND chips in a pure software way: the flash management algorithms are directly implemented by the OS, mainly through the use of dedicated Flash File Systems (FFS). Linux implements the most popular FFS [12–14].

Figure 1 depicts the embedded Linux software stack for NAND secondary storage management. User space programs (A in Figure 1) access files using system calls (open, read, etc.). These calls are received by the Virtual File System (VFS, B) layer. VFS is used to abstract the underlying file systems specificities and to present them in a unified way to the user. VFS relies on the actual file system, which in our case is a FFS (C). To access the raw flash chip (E), the FFS uses a NAND driver. In embedded Linux, this driver is implemented inside the *Memory Technology Device* (MTD) software layer (D).

The MTD subsystem [8, 15] roles are to provide a unified access to various semiconductor storage memories such as flash, RAM / ROM, etc ; and to implement drivers for all those memory devices. Within the scope of our study, MTD can be seen as a generic abstraction layer and drivers container for all the models of raw flash chips supported by Linux. MTD provides various ways to access NAND chips: 1) with a unified API exported in the kernel, 2) as character devices, 3) as block devices. MTD also provides partitioning support.

MTD can be seen as a software stack. The layers of this stack are functions involved in flash I/O management, calling one another in a top-down way. From the MTD data structures point of view, the functions composing the MTD stack are stored in function pointers. This allows the different flash chip designers to replace MTD default functions by their own driver primitives. The upper part of this stack represents abstraction and translation layers related functions. These generic flash access functions are called by the FFS layer. The bottom of this stack is the driver layer, containing specific and platform dependant functions for manipulating each of the supported flash chips. As the upper part of the stack is constituted of generic functions, the bottom functions are platform specific.

## 3. FLASHMON VERSION 2: IMPLEMENTATION AND FEATURES

In this section we give details on the implementation and usage of Flashmon. The upgrades brought by the version 2 (v2) from version 1 (v1) are also outlined.

### 3.1 Core concepts

Flashmon (F in Figure 1) is a Linux kernel module monitoring function calls at the MTD level to trace flash operations. It can be loaded and unloaded dynamically at runtime. It uses kernel probes [7], more particularly *Jprobes*, placed on the MTD functions corresponding to the page read (G in Figure 1), page write (H) and block erase (I) NAND operations. Kernel probes are data objects linked to a kernel function to trace. A handler in Flashmon is associated with each of the probes and is executed each time the probed function is called. In the handler, Jprobes provide access to the parameters values of the probed function. This allows Flashmon to determine the address of the corresponding operation and then to log the event type, address and arrival time in RAM. In fact, all the Flashmon data structures are kept in RAM in order to keep the tool as less intrusive as possible and thus reducing the interference with the traced I/O operations. The trace log kept in RAM is available for output in various formats which will be discussed further in the paper. Tracing at the MTD level allows Flashmon to work with different FFS, and to support several models of flash memory chips.

### 3.2 Probed Functions

In Flashmon v1, probed functions were the generic *nand_read*, *nand_write* and *nand_erase* functions. They are located in the upper part of the MTD stack. This allowed the tool to work on many platforms because these are very high level MTD functions that are directly called by the FFS. A major drawback of this solution is that sometimes several pages are read or written through one call of these functions. Time measurements in the handler for the last page of the read / written set were inaccurate when the set was important, because the handler is entirely executed before the set of flash memory operations is launched. V2 solves this problem by tracing lower level functions corresponding to one unique page read / write operation.

Determining the right functions to probe is not a trivial problem: they have to be as low level as possible to be the closest to the actual hardware traced flash operation. They also have to be generic enough to be called with each of the NAND chip models supported by Linux.

In fact, in v2 the probed function names are not hard coded in the module. At launch time, Flashmon performs a search in MTD data structures to determine the optimal function to trace. A sub-module of Flashmon called the "function finder" is in charge of this work. This module follows the function pointers of the MTD stack presented earlier to obtain the addresses of low level but still generic functions to probe. Moreover, according to the kernel version, it may fall back on higher level functions. Indeed, in earlier Linux versions, low level MTD functions do not have enough information in their parameters for Flashmon to perform a precise trace. The function finder sub-module is written in a very generic way allowing easy extension of Flashmon to support new flash models and kernel code changes.

### 3.3 Outputs

Flashmon offers two outputs in the /proc virtual file system: the *spatial view* /proc/flashmon and the *temporal log* /proc/flashmon_log.

The spatial view (K on Figure 1) file contains a number of lines equal to the number of blocks in the traced flash chip. Each line contains 3 values representing respectively the number of page reads, page writes, and block erasures sustained by a block. The file is built on demand each time it is read. Spatial view is useful to see the flash state at a given time and in particular to observe the distribution of erase operations to evaluate the wear leveling.

The temporal log (J on Figure 1) is a novelty of v2. It is also built on demand when it is read. Each line of the file contains four coma separated fields corresponding to one logged event: the arrival time of the event, the event type (page read, page write, block erase), the targeted address (index of page read / written and block erased), and the current process executed when the operation was traced. One example of such a file is as follows:

```
13.551048336;R;22655;cat
13.552904998;W;6935;sync_supers
13.563917567;E;1025;jffs2_gcd_mtd6
```

To build this log, Flashmon relies on internal RAM data structures. One of these structures is populated each time a flash event is traced. They are all stored in a buffer whose maximum size is configurable (the entire log is allocated at launch time): as the number of logged events may become important on intensive workloads, it gives the user a way to control Flashmon RAM usage. The buffer is a circular log buffer: when it becomes full, older entries are overwritten with newer ones. The logged time for each event is acquired with the *getnstimeofday()* system call with a nanosecond precision. The location where Flashmon's trace is gathered, MTD, is fully synchronous. Therefore, Flashmon is not disturbed by concurrent accesses generated in a multi-threaded environnement.

### 3.4 Others Features

Flashmon provides various complementary features to ease and customize the tracing process: 1) single partition tracing, 2) user space notification, 3) tracing process control, 4) complementary tools and kernel source integration.

As Flashmon v1 allowed only to trace the entire chip, Flashmon v2 allows tracing only one partition if needed. Launching the v1 module required several mandatory parameters such as page and block sizes. Flashmon v2 examines the MTD data structures at launch time and collects itself the needed information for the traced chip or partition: it can be launched without parameters for a fast and simple monitoring.

When Flashmon is inserted, one can customize the tracing process by providing optional parameters which are: (1) the index of a partition to trace; (2) the maximum size for the temporal log and (3) the PID of a user space process to notify each time a flash access is traced. The notification is a feature from v1. If selected, Flashmon will send a signal to the user space process each time a flash access is traced. It avoids active standby for processes monitoring the spatial view. Flashmon monitoring can be controlled by writing commands to /proc/flashmon and /proc/flashmon_log to stop, pause or reset the tracing process, and to flush the temporal log.

Version 2 also comes with a set of tools to ease Flashmon usage and format its outputs. A couple of shell scripts allow plotting the outputs of spatial view and temporal log. The results presented in the case study section are obtained through the use of these scripts. Another script is provided to patch an existing kernel source directory and integrate Flashmon in these sources. Flashmon can then be selected as a module or as a built-in feature using the kernel compilation configuration menu. One benefit for selecting Flashmon as a built-in function is the fact that the tracer is loaded before the file system driver at boot time, this allows to trace flash operations during the kernel boot process. Results on that topic are provided in the case study section.

Flashmon source code is about 800 lines of C. It comes with a complete and up-to-date documentation. The provided Makefiles should allow (cross-)compiling the module for most of the platforms. Concerning Flashmon dependencies, the module should be compiled against a kernel with the kprobes feature enabled, and MTD NAND support.

### 4. CASE STUDIES

In this section we present results obtained with Flashmon when tracing flash operations during (1) the kernel boot process and (2) the execution of the Postmark [6] benchmark. These experimentations were obtained on the Armadeus APF27 development board [1], embedding an ARM9 based Freescale I.MX27 CPU clocked at 400 MHz and 128 MB of RAM. The board is in particular equipped with a 256 MB Micron SLC NAND flash chip [9] containing one die, one plane, and 2048 blocks of 64 pages each. The page size is 2KB.

### 4.1 The Kernel Boot Process

Flashmon was used to trace NAND operations during the boot process, with a root file system (rootfs) stored on a flash partition. During that process, secondary storage I/O accesses begin when the file system driver is loaded and the rootfs mounted.

#### 4.1.1 Methodology

We included Flashmon as a built-in feature in the 2.6.38 Linux kernel version obtained with the Armadeus toolchain. The Armadeus default kernel configuration was used, with the addition of the kernel probes feature activation. A standard embedded rootfs was flashed on an erased 50 MB partition of the

NAND chip. The kernel was launched and Flashmon results were collected when the boot process was finished. The system was then rebooted and a new set of results were collected for this second boot. Two consecutive boot processes were traced to observe the differences between the first boot after flashing a new rootfs, and a more standard boot after a system shutdown. Indeed, FFS perform a formatting operation during the first mount operation. Experiments were launched on two file systems' rootfs: JFFS2 and UBIFS.

### 4.1.2 Results and Discussion

Results are depicted in Figure 2. The rootfs partition goes from page 4096 to page 29696. Before the experimentation, the bootloader flashes sequentially the rootfs starting from page 4096 on the erased rootfs flash partition. The rootfs size is 7.5 MB for JFFS2 and 10 MB for UBIFS, so the last page containing actual rootfs data for the first boot is page 7936 for JFFS2 and page 9216 for UBIFS.

For each FFS and boot process we can observe several phases. First, when the file system is mounted the entire rootfs partition is scanned (A on Figure 2), represented by a succession of sequential read requests. That scan takes considerably more time with JFFS2. One can see in Flashmon trace that JFFS2 reads each page of the partition while UBIFS scans only the first page of each block (not visible on the plot because of the scale). Next, comes a set of mainly read operations (B in Figure 2). It is the */etc/rcS* boot script which loads a set of scripts to initialize various services (network interfaces, ssh server, etc.). Flash accesses consist mostly in reading and writing configuration files for these services. After some time, the login prompt is available on the serial output, represented by the double-headed arrow on the x axis.

JFFS2 mount process consists of a two phase scan. The partition is first fully scanned (A), then a kernel thread performs in the background a meticulous scan of all existing file data to check the file system consistency (CRC scan). This background process starting just after the end of the first scan up to 65 seconds can seriously degrade performance of I/O requests occurring during its execution. UBIFS mount operation is done considerably faster than JFFS2.

For the first boot process, a formatting phase can be observed in both FFS (C). The partition blocks that do not contain data are erased sequentially. This is done just after the first scan with UBIFS, and with the background kernel thread after the login prompt for JFFS2. It is interesting to note that even though the partition was fully erased before the flash of the rootfs (in the bootloader), the FFS still re-erases the blocks that do not contain data. Differences in the logging prompt appearance time between the first and the second boot are mainly due to the generation of keys by the SSH server, done only during the first boot process.

## 4.2 Postmark Benchmark

Postmark [6] is a synthetic macro benchmark designed to reproduce the behavior of a file system containing many short lived small files under a heavy load. One execution of Postmark consists in several phases. First, a set of initial files are created, distributed in subdirectories. In the next phase, transactions are performed on those files. A transaction consists of two operations: the creation or deletion of a random file, then a read or append operation on another random file. Once the transaction phase is complete, all the created files are deleted. As one could argue that Postmark is not a specifically embedded FS benchmark, we think that it is generic enough to reproduce the behavior of many of today's embedded applications.

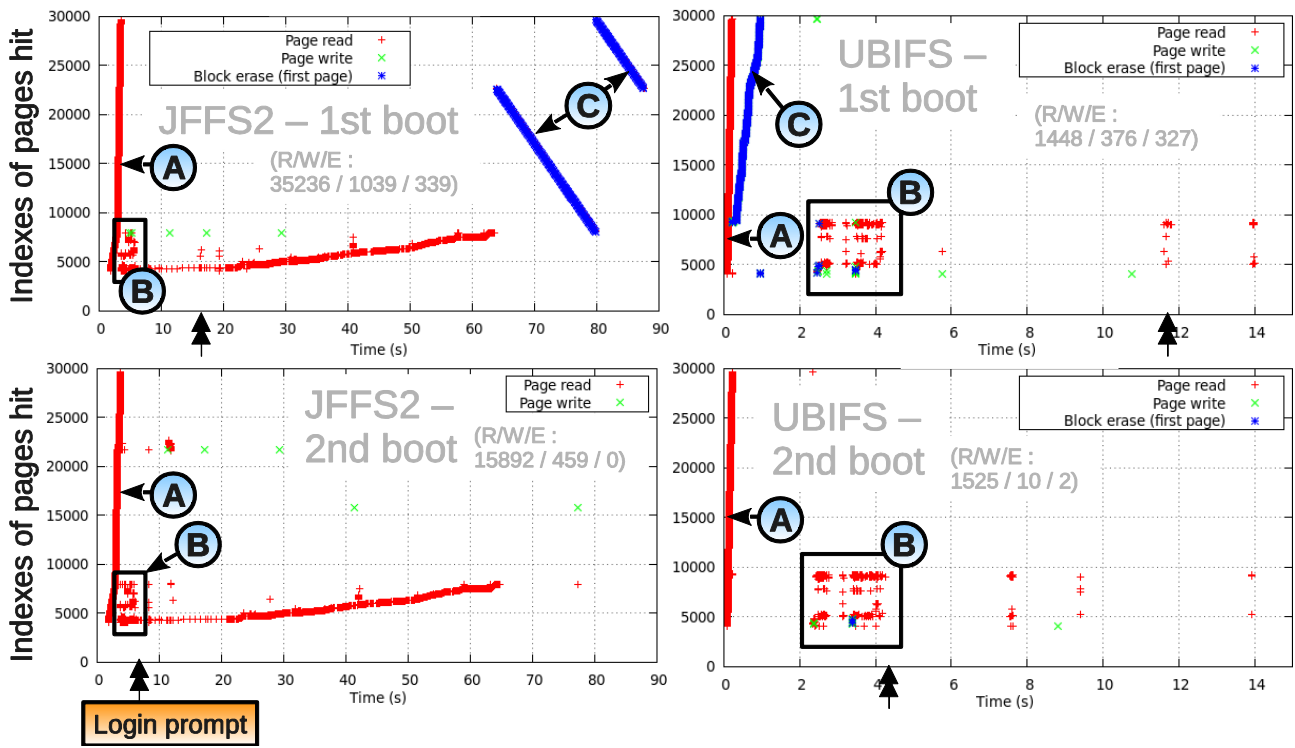

**Figure 2: NAND operations during the Linux kernel boot process. The x axis range is different for JFFS2 and UBIFS.**

### 4.2.1 Methodology

The Postmark configuration presented in Table 1 was used. Created files and transactions counters parameters have to be large enough to generate a significant number of flash operations: A too small workload would be in majority absorbed by the Linux page cache. Using synchronous I/O or bypassing the page cache through the O_SYNC or O_DIRECT open / mount flags is not an option because the FFS layer does not support such features. Note that most of the FFS compress data before writing on the flash media so one cannot guarantee that a 4 KB high level read / write request will end up as a 4 KB flash read / write request.

**Table 1. Used Postmark configuration**

| Parameter | Value |
| --- | --- |
| Number of files created initially | 800 |
| Created files size | Between 512 bytes and 10 KB |
| Number of transactions | 3000 |
| Size of all read and write requests | 4 KB |
| Transaction read / append ratio | 50 % |
| Transaction create / delete ratio | 50 % |
| Number of subdirectories | 10 |

Postmark was launched on a clean 50 MB dedicated flash partition. The experimentation was repeated on a JFFS2, YAFFS2 and UBIFS partition. The kernel was patched to support YAFFS2. After the end of the benchmark we waited several seconds because of potential asynchronous garbage collection processes, and then dumped the Flashmon temporal log. Postmark reports for this configuration a total of 9.25 MB read and 14.45 MB written data volume.

### 4.2.2 Results

Results are depicted on Figure 3. The Y axis represents the entire address space of the test partition. For each of the FFS we can clearly observe the creation phase (A on Figure 3), which consists in a majority of write requests. The fact that the partition is clean allows FFS to perform sequential writes. The transaction phase comes next (B), as a set of read and write requests. Writes (file creations and updates) continue to be performed sequentially, illustrating the out-of-place updates feature of each FFS. During this transaction phase, read requests are also performed: they correspond to 1) regular read requests and 2) read requests performed by the file system to gather data / metadata in order to satisfy write requests. Read are performed on previously written data.

At the end of the transaction phase (end of B), the Postmark process returns after having deleted all the created files. We can then observe for JFFS2 and YAFFS2 an asynchronous garbage collection (C) phase, which consists in erasing the blocks containing data invalidated by the files deletion. The garbage collector (GC) is generally implemented in the form of a kernel thread. GC execution is based on various thresholds such as the quantity of clean space available, and the amount of invalid data. According to these thresholds the YAFFS2 GC can be launched with more or less "aggressiveness": we can clearly observe an aggressive GC phase (C1) followed by a soft GC phase (C2). Note that YAFFS2 soft GC continues for 7 minutes after the ending of Postmark. Finally, we can see that UBIFS does not perform any GC, because of the low quantity of data written during the benchmark. UBIFS is a strongly buffered FFS as compared to JFFS2 and YAFFS2 that are more synchronous. UBIFS buffers absorb a large amount of the Postmark workload. JFFS2 and especially YAFFS2 write a lot more data, and it impacts strongly the execution time of Postmark. The number of flash operations performed by each FFS is also present on Figure 3. We can see that YAFFS2 performs twice the number of page writes and block erasures as JFFS2. UBIFS performs 20 times less write operations than YAFFS2.

Those results show that Flashmon can supplement performance evaluation, based on execution times, with essential knowledge concerning the actual flash operations occurring during the tests: their types, numbers, and distribution. It helps in explaining execution time related results that can be affected by asynchronous processes, caches, etc. The impact of such elements can be hard to detect and characterize without tracing tools.

## 5. AN ANALYSIS OF FLASHMON IMPACT ON THE TRACED SYSTEM

In the monitored system, Flashmon is not intrusive in terms of number of flash I/O operations performed because it keeps all its internal structures in RAM. In order to give a realistic trace, the overhead of Flashmon on I/O performance must be minimal. Moreover, as it is primarily designed to be used in embedded systems, the RAM footprint of the tool must be controllable.

### 5.1 Impact on I/O Latency

To study Flashmon impact on I/O performance we launched several I/O intensive experimentations without and with Flashmon. Then we observed the execution time differences.

A first set of experimentations consisted in erasing a 100 MB flash partition, write a 5 MB file containing random data in this partition then read this file. Those operations were done through the use of the MTD utilities *flash_erase*, *nandwrite* and *nanddump*, [8] which allow bypassing the file system layer and address directly physical flash pages and blocks. Each operation was launched with and without Flashmon several times, execution times were measured with the *gettimeofday()* system call, and mean values were computed.

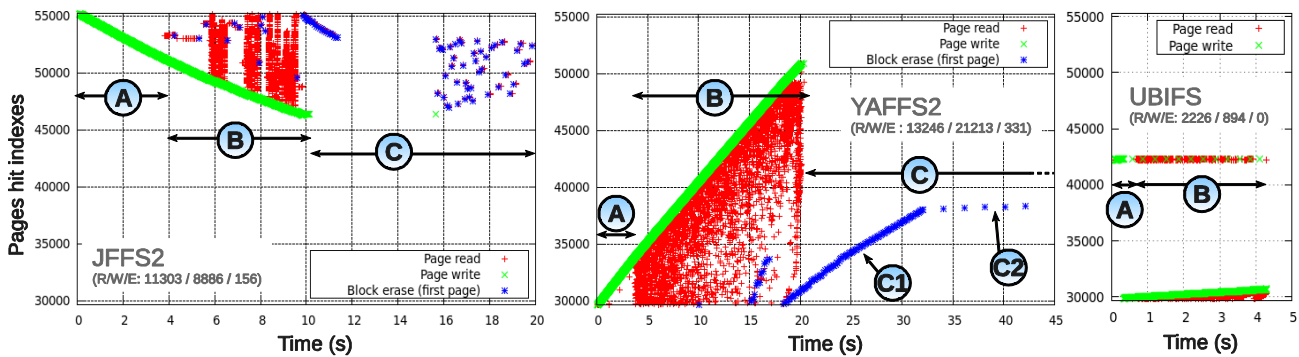

**Figure 3: Postmark results – The total number of read / write / erase operations is indicated under the title of each picture.**

Another set of experimentations consisted in launching the Postmark configuration presented in the previous section (See Table 1) and measuring the execution time. Once again, mean values from several executions, with and without Flashmon, were used to compute the tool overhead. Those experimentations were launched on JFFS2, YAFFS2 and UBIFS. Results from both experimentation sets are presented on Table 2. All those results show that Flashmon overhead on I/O performance in the traced system stays under 6%.

**Table 2. Flashmon performance impact**

| First set of experimentations | | | |
|---|---|---|---|
| **Experiment** | **Mean execution time (s)** | | **Overhead (%)** |
| | **Without Flashmon** | **With Flashmon** | |
| Erase 100 MB | 0.470 | 0.489 | *3.85* |
| Write 5 MB | 1.498 | 1.570 | *4.79* |
| Read 5 MB | 2.386 | 2.515 | *5.40* |
| *Second set of experimentations* | | | |
| **FFS** | **Postmark mean exec. time (s)** | | **Overhead (%)** |
| | **Without Flashmon** | **With Flashmon** | |
| JFFS2 | 10,523589 | 10,628663 | *3,08* |
| YAFFS2 | 19,793678 | 20,626676 | *4,21* |
| UBIFS | 4,690151 | 4,754635 | *1,37* |

## 5.2 Flashmon Memory Footprint

One can model Flashmon memory usage as the sum of the static and dynamic parts of the memory used by the module. The static part is all the statically allocated memory. It can be obtained by observing the Flashmon entry in */proc/modules*. For the current version it is 8861 bytes. The dynamic part consists in memory allocated with *vmalloc*() (the *malloc*() kernel equivalent).

The dynamic part is divided into (1) the counters for the spatial view and (2) the temporal log data structures. For (1), there is one set of three counters for each of the flash blocks of the traced partition / chip, one counter for each type of operation (R/W/E). Each counter is an unsigned 32 bits integer. Regarding (2), for each flash event inserted in the temporal log a data object is inserted in a buffer which maximum size is configurable at launch time. The entire buffer is allocated when Flashmon is inserted. One can obtain the size of one entry with *sizeof()*, for the current version it is 20 bytes + 16 bytes to store the task name for each entry.

Flashmon memory usage can then be modeled as follows: $static+dynamic = 8861 + 3*4*N_{FlashBlocks} + 36*MaxLogSize$ bytes

$N_{FlashBlocks}$ is the number of flash blocks in the traced partition and *MaxLogSize* the maximum number of entries present in the temporal log. Flashmon memory usage is primarily impacted by the number of log entries. The configuration of Flashmon used in the case studies was the following: with 2048 blocks traced and a max log size of 40000 elements, Flashmon uses 1.4 MB of RAM. The fact that the current task name is logged with each event impacts greatly the memory usage, so an option is provided to disable this feature. A solution to reduce memory usage is also to limit the maximum log size and to flush regularly the results in a non intrusive way, for example over the network.

## 6. CONCLUSION

In this paper, Flashmon version 2 is presented. This Linux kernel module monitors at runtime NAND flash operations on embedded raw flash chips. Flashmon employs kernel probes. It can be used to study today's flash management mechanisms behavior and performance, and propose optimizations. Flashmon allows completing performance evaluations based on execution time with essential qualitative knowledge on events occurring in the flash layer. It can also help in prototyping and validating new flash management systems in real conditions. Flashmon is generic as it is designed to be used with all Flash File Systems. It is very flexible as it allows to easily customize and control the tracing process. Flashmon is non intrusive and has a low overhead on the performance (under 6% on the tested workloads) of the traced system. Its memory footprint is also configurable. Flashmon comes with a complete documentation and is available under the GPL license: http://sourceforge.net/projects/flashmon/.